\documentclass[runningheads]{llncs}
\usepackage[T1]{fontenc}
\usepackage{graphicx}
\usepackage{array}
\usepackage{tabularx}
\usepackage{hyperref}
\usepackage{amsmath}
\usepackage{tikz}
\usepackage{algorithm}
\usepackage{algpseudocode}
\usetikzlibrary{shapes.geometric, arrows.meta, positioning}

\usepackage{color}

\begin{document}
\title{A proposal for PU classification under Non-SCAR using clustering and logistic model}
\titlerunning{A proposal for PU ...}
%
\author{Konrad Furma\'{n}czyk\inst{1}\orcidID{0000-0002-7683-4787} \and
Kacper Paczutkowski\inst{1}\orcidID{0000-0001-7408-6060}}

\authorrunning{Furmańczyk K. et al.}
\institute{Institute of Information Technology, Warsaw University of Life Sciences, Warsaw, Poland
\email{\{konrad\_furmanczyk,kacper\_paczutkowski\}@sggw.edu.pl}}

\maketitle              

\begin{abstract}
The present study aims to investigate a cluster cleaning algorithm that is both computationally simple and capable of solving the PU classification when the SCAR condition is unsatisfied. A secondary objective of this study is to determine the robustness of the LassoJoint method to perturbations of the SCAR condition. In the first step of our algorithm, we obtain cleaning labels from 2-means clustering. Subsequently, we perform logistic regression on the cleaned data, assigning positive labels from the cleaning algorithm with additional true positive observations. The remaining observations are assigned the negative label. The proposed algorithm is evaluated by comparing 11 real data sets from machine learning repositories and a synthetic set. The findings obtained from this study demonstrate the efficacy of the clustering algorithm in scenarios where the SCAR condition is violated and further underscore the moderate robustness of the LassoJoint algorithm in this context.

\end{abstract}
\section{Introduction}
Learning from positive and unlabeled data (PU) is an approach in which a dataset has only positive examples and unlabeled data. In the PU setting, the training data contains positive and unlabeled examples, which means that the true labels $Y\in \left\{ 0,1\right\} $ are not observed directly in the data. We only observe the surrogate variable $S\in \left\{ 0,1\right\}$, which indicates whether an example is labeled (and consequently positive, $S=1$) or not ($S=0$). For example, consider the problem of predicting disease based on patient characteristics. Some patients are properly diagnosed (the surrogate variable $S=1$ and consequently, the true label $Y=1$ in this case), there also exists a group of patients without diagnosed disease (we put $S=0$ then). The second group includes both patients who have not actually been diagnosed with the disease but have the disease (in this case, we have Y = 1 and S = 0) and patients who do not actually have the disease (in this case, we have Y = 0 and S = 0). 
PU learning has been successfully applied in vision tasks \cite{L1}, text classifications \cite{L2,L3}, recommendation systems \cite{Yi} and medical diagnosis predictions \cite{C,R}.
In particular, the PU learning methods can be applied in the case when under-reporting is present in survey data \cite{Bek}. Some other interesting examples where the under-reporting is present may be found in the paper by Bekker and Davis \cite{Bek}. Now, suppose that $X$ is a feature vector and, as previously, $Y\in\left\{ 0,1\right\}$ stands for a true class label and $S\in \left\{0,1\right\}$ denotes the surrogate variable that indicates whether an example is labeled ($S=1$) or not ($S=0$). We consider a single sample scenario, where it is assumed that there is a certain unknown distribution $P$ of $(Y,X,S)$, such that $(Y_i,X_i,S_i), i=1,\ldots,n,$ form an iid sample from this distribution. We do not have a traditional sample $(X_i,Y_i)$, which is considered in standard classification problems. We only observe a sample $(X_i,S_i)$, where $S_i$ are the observations of variable $S\in \left\{ 0,1\right\}$  (since $S$ is a surrogate of the true label $Y$, then each $S_i$ depends on $(X_i,Y_i)$). In the considered concept only positive examples (i.e., examples for which $Y=1$) may be labeled, which means that $P(S=1|X,Y=0)=0.$ It should be emphasized that in the PU design, the true class labels $Y$ are only partially observed, which means that if $S=1$, then we know that $Y=1$, but if $S=0$, then $Y$ may be either $1$ or $0$.
\newline
First, we consider the Selected Completely At Random (SCAR) condition, where is assumed $$P(S=1|Y=1,X)=P(S=1|Y=1).$$
The SCAR assumption implies that $X$ and $S$ are independent given $Y,$ since $P(S=1|Y=0,X)=P(S=1|Y=0)=0.$
 Let $c=P(S=1|Y=1).$ The parameter $c$ is called the label frequency and plays a key role in the PU learning scheme. The SCAR assumption significantly simplifies the estimation of the posterior probability $P(Y=1|X=x)$, where, as previously, $Y \in \{0,1\}$ denotes a true class label and $X$ stands for the feature vector. Under this assumption, we have $$P(Y=1|X=x)=c^{-1} P(S=1|X=x).$$ The label frequency $c$ can be estimated using the observed PU data \cite{Laz}.
 In many practical applications the SCAR assumption does not hold. For example, some factors $X$ may affect the label frequency process. In this paper, we consider the SAR (Selected at Random) assumption, which means $$P(S=1|Y=1,X=x)=e(x),$$ where $e$ is some unknown function of feature vector $X=x$. The quantity $e(x)$ is called the propensity score. 
 \\Recently, a few methods were invented for the SAR condition, especially based on the Expectation-Maximization (EM) algorithm \cite{Bek1,Gong} and using joint maximum likelihood for the double logistic model for $e(x)$ and $y(x)=P(Y=1|X=x)$ \cite{F4}. 
\\Based on the logistic model, three basic methods of this estimation have been proposed so far. They minimize the empirical risk of the logistic loss function and are known as the naive method, the weighted method, and the joint method (the latter has been introduced in Teisseyre et. al. \cite{Tes}). In \cite{F1}-\cite{F3} was proposed a new method under the SCAR assumption called the LassoJoint. This method is based on the logistic model (joint method) and thresholded Lasso selection. The present study focuses on the logistic regression approach, which is closely associated with the concept of easy interoperability. However, for the problem of PU, generative adversarial networks (GANs) have also been employed in the works of Hou et al. \cite{Hu}  and Guo et al. \cite{Gu}.
\\The objective of this study is to investigate a computationally efficient cluster cleaning algorithm that can address the PU problem when the SCAR condition is not satisfied. The second goal of the paper is to check how the LassoJoint method is robust to such a perturbation of the SCAR condition.
The remainder of our paper is structured as follows. In Section 2 we introduce our aims and methods, in Section 3 we present our numerical experiments together with the obtained results, in Section 4 we summarize and conclude our study. Comprehensive results of our study are available in Supplement \footnote{https://github.com/kapacc/mdai25-clusters-scripts}.
This supplement also includes all codes in R that are fully reproducible.
In order to execute our simulations, we employed the RStudio server module from the ICM UW Topola server\footnote{This research was carried out with the support of the Interdisciplinary Centre for Mathematical and Computational Modelling (ICM) at the University of Warsaw, under computational allocation No. g99-2175.}. We applied the following libraries:  glmnet \cite{Fr,glmnet},  caret \cite{Caret}, stringr \cite{Stringr}, dplyr \cite{Dplyr}, pROC \cite{pROC}, and MLmetrics \cite{MLmetrics}.

\section{Objectives and Methods}
In this section, we describe our methods and objectives. First, we present the LassoJoint method and the naive method. Next, we show our proposed cluster cleaning method. Finally, we recall the basic classification metrics that we will use.

Our main goal is to investigate a computationally efficient cluster cleaning algorithm that can be helpful in solving the PU classification problem when the SCAR condition is not satisfied. The second objective is to determine the resilience of the LassoJoint method to a perturbation of the SCAR condition.
\\The LassoJoint method \cite{F1} can be described as follows:
(1) For available PU dataset $(S_{i},X_{i})$, $i=1,\ldots ,n$, we perform
the ordinary Lasso procedure (see Tibshirani \cite{T}) for some tuning
parameter $\lambda >0$, i.e. we compute the following Lasso estimator of $%
\beta ^{\ast }=-argmin_{\beta}E_{X,Y}l(\beta,X,Y)$ and $l$ is the logistic loss (see \cite{F3} for more details)
$$\widehat{\beta}^{(L)}=\arg \min_{\beta\in R^{p+1}}\widehat{R}(\beta )+\lambda
\sum_{j=1}^{p}\left\vert \beta _{j}\right\vert,$$
where $$\widehat{R}(\beta )=-\frac{1}{n}\sum_{i=1}^{n}\left[ S_{i}\log
\left( \sigma (X_{i}^{T}\beta )\right) +\left( 1-S_{i}\right) \log \left(
1-\sigma (X_{i}^{T}\beta )\right) \right],$$ where $\sigma(t)=exp(t)/(1+exp(t))$
and subsequently, we obtain the corresponding support $Supp^{(L)}=\{1\leq
j\leq p:\widehat{\beta}_{j}^{(L)}\neq 0\}$;
(2) We perform the thresholded Lasso for some prespecified level $\delta $
and obtain the support $Supp^{(TL)}=\{1\leq j\leq p:\left\vert \widehat{\beta}%
_{j}^{(L)}\right\vert \geq \delta \}$;
(3) We apply the joint method from Teisseyre et al. \cite{Tes} for the
predictors from $Supp^{(TL)}$.

It should be stressed that under some mild regularity conditions, the LassoJoint procedure obeys the screening property (all significant predictors of the model are chosen in the first two steps, see Theorem 1(b) in \cite{F1}). This method includes a selection step and can be applied to both high-dimensional and low-dimensional data.
\newline
\\The naive method is when we applied logistic regression for $S=1$ and $S=0$ naive considering surrogate labels $S$ as true labels for $Y$.
\subsection{Cluster Cleaning Procedure}
We propose a new cluster cleaning method. In the first step, this method produces two clusters from data where the proxy label is $S=0$. Observations from the cluster with more observations with label $S=1$ than with label $S=0$ are assigned the label $\hat{Y}=1$, and observations from the other cluster are assigned the label $\hat{Y}=0$. This is a cleaning step for our algorithm. Next, we perform logistic regression on the cleaned data, assigning $\hat{Y}=1$ from the cleaning algorithm with added observations that are true $Y=1$ as $S=1$, and we consider the rest of the observations as $\hat{Y}=0$.
\\Now, we will describe our procedure in more detail. We named this procedure "pecking" because it resembles pecking out a certain part of the data.
First, we draw ("peck") $q\%$ observations with layer $S=1$ and add these drawn observations to layer $S=0$. This creates the set $C_0^q$. 
Subsequently, the 2-means clustering algorithm is applied to observations from $C_0^q$ in order to obtain two clusters: $C_{0,0}$ and $C_{0,1}$, such that $C_0^q = C_{0,0} \cup C_{0,1}$.
A cluster with more labels, $S=1$, is denoted as $C_{0,1}$. 
Finally, two disjoint sets are identified, where we assign label $\hat{Y}=1$ for layer $\{S=1\} \cup C_{0,1}$ and label $\hat{Y}=0$ for the rest of observations $C_{0,0}$. 
We repeat this procedure $R$ times. 
\\Next, for each repetition, we fitted logistic regression for data with cleaned labels $(\hat{Y}_i,X_i), i=1,\ldots,n$ from the first scenario. In the second scenario, for each repetition, we apply the LassoJoint procedure with two variants:  strict(averages coefficients for features that appear in all repetitions) and non-strict(averages coefficients for features that appear in any repetition). The average values of the parameters for all repetitions are taken to derive the final predictive model. (In other words, we repeat pecking the entire set of $Y=1$ in $R$ times and average the obtained GLM coefficients). The pseudocode for this algorithm is given below.
\\
\begin{algorithm}[H]
\caption{Pecking Procedure for PU Learning}
\begin{algorithmic}[1]
\Require Dataset $(X_i, S_i)$ for $i = 1, \ldots, n$ with $S_i \in \{0,1\}$, proportion $q \in (0,1)$, repetitions $R$
\Ensure Final predictive model coefficients

\For{$r = 1$ to $R$}
    \State Draw $q\%$ of samples with $S=1$ and add to samples with $S=0$ to form $C_0^q$
    \State Apply 2-means clustering on $C_0^q$ to get clusters $C_{0,0}$ and $C_{0,1}$

    \If{$C_{0,0}$ has more samples with $S=1$ than $S=0$}
        \State Swap $C_{0,0} \leftrightarrow C_{0,1}$ \Comment{Ensure $C_{0,1}$ has more $S=1$}
    \EndIf

    \State Assign labels:
    \State \quad $\hat{Y} = 1$ for all samples in $\{S=1\} \cup C_{0,1}$
    \State \quad $\hat{Y} = 0$ for all samples in $C_{0,0} \setminus \{S=1\}$

    \State Fit either:
    \State \quad Logistic regression on $(X_i, \hat{Y}_i)$
    \State \quad or LassoJoint (strict / non-strict variant)

    \State Store resulting coefficients $\beta^{(r)}$
\EndFor

\State Compute final coefficients:
\If{Logistic regression}
    \State Average all $\beta_j$ across $R$ repetitions
\ElsIf{LassoJoint (strict)}
    \State Average $\beta_j$ only for features appearing in \textbf{all} $R$ repetitions
\Else \Comment{LassoJoint non-strict}
    \State Average $\beta_j$ for features appearing in \textbf{any} repetition
\EndIf

\end{algorithmic}
\end{algorithm}
To date, the PU research literature has employed a few approaches based on clustering methods. For instance, in \cite{LX},
in the context of text classification, the initial step of this approach involves the collection of reliable negative examples (C-CRNE) through hierarchical clustering. 
During the process of building the classifier, the Term Frequency Inverse Positive-Negative Document Frequency approach is adopted.
In contrast to the aforementioned approach, the proposed method does not exclusively focus on text classification. 
It is also computationally more efficient and expeditious due to its utilisation of the 2-means method in place of hierarchical clustering.

\subsection{Classification Metrics}
The accuracy and F1-score are defined as follows:
$$Accuracy=\frac{TP+TN}{TP+FP+FN+TN},$$
$$F1 = \frac{2 \cdot Precision \cdot Recall }{Precision + Recall},$$ 
where $TP,$ $FN,$ $TN$ and $FP$ stand for: the number of true positives, false negatives, true negatives and false positives, respectively. We will also report the AUC value, a summary metric of the ROC curve that reflects the test's ability to distinguish between positive and negative individuals.
\section{Numerical Experiments}
\subsection{Datasets}
The assessments of the mentioned metrics for the naive method, the clust method (in two scenarios) have been gained by conducting some numerical study on one synthetic and 11 real low-dimensional datasets from the UCI Machine Learning Repository \cite{uci}. Our synthetic dataset (Artif) was created by generating a matrix \( X \) with both relevant and irrelevant features, where the first few variables have increasing standard deviations. A logistic function was then applied to these features to probabilistically generate the binary target variable \( Y \), based on the influence of the relevant features.
We filtered out quasi-constant features where the most common value constituted more than 90\% of the observations and removed highly correlated features using a correlation threshold to reduce redundancy. Numeric features were scaled using Min-Max scaling, and only those with at least 5 unique values were retained to ensure continuous features were preserved for analysis. The summary table provides key characteristics of various datasets, including the number of features, observations, non-continuous variables, continuous variables, negative and positive instances, the percentage of positive instances, and the mean absolute correlation, defined as follows:

\[
\text{Mean Abs Corr} = \frac{2}{p(p-1)} \sum_{i=1}^{p-1} \sum_{j=i+1}^{p} \left| \rho_{ij} \right|
\]

where:
\begin{itemize}
  \item \( \rho_{ij} \) is the Pearson correlation coefficient between feature \( i \) and feature \( j \).
  \item \( p \) is the total number of features (excluding the target variable \( Y \)).
\end{itemize}
As seen in Table \ref{tab:summary_table}, the selected datasets are diverse in terms of size and complexity. In the context of our summary table, we considered variables as non-continuous if they have fewer than 15 unique values.

\begin{table}[ht]
\centering
\begin{tabularx}{\textwidth}{|l|X|X|X|X|X|X|X|X|}
\hline
\textbf{Dataset} & \textbf{Feat} & \textbf{Obs} & \textbf{Non-Cont Var} & \textbf{Cont Var} & \textbf{Neg Inst} & \textbf{Pos Inst} & \textbf{Pos \%} & \textbf{Mean Abs Corr} \\ \hline
adult & 5 & 32561 & 0 & 5 & 7841 & 24720 & 75.92 & 0.07 \\ \hline
artif & 21 & 2000 & 0 & 21 & 977 & 1023 & 51.15 & 0.02 \\ \hline
bank-marketing & 10 & 4119 & 3 & 7 & 451 & 3668 & 89.05 & 0.23 \\ \hline
banknote & 5 & 1372 & 0 & 5 & 610 & 762 & 55.54 & 0.43 \\ \hline
breastc & 10 & 683 & 9 & 1 & 239 & 444 & 65.01 & 0.6 \\ \hline
credit-a & 7 & 653 & 0 & 7 & 296 & 357 & 54.67 & 0.17 \\ \hline
credit-g & 7 & 1000 & 3 & 4 & 300 & 700 & 70 & 0.12 \\ \hline
dhfr & 107 & 325 & 16 & 91 & 122 & 203 & 62.46 & 0.25 \\ \hline
diabetes & 9 & 768 & 0 & 9 & 268 & 500 & 65.1 & 0.17 \\ \hline
spambase & 39 & 4601 & 0 & 39 & 1813 & 2788 & 60.6 & 0.07 \\ \hline
wdbc & 29 & 569 & 0 & 29 & 212 & 357 & 62.74 & 0.38 \\ \hline
wine-quality & 12 & 1599 & 0 & 12 & 217 & 1382 & 86.43 & 0.2 \\ \hline

\end{tabularx}
\caption{Summary of Dataset Properties}
\label{tab:summary_table}
\end{table}

\subsection{Non-SCAR Scenarios}
In the SCAR (Selection Completely At Random) scheme, it is easy to specify the coefficient \( c \) in advance when generating the surrogate variable \( S \), as the selection process is completely random and independent of the variables' values. However, in the non-SCAR scheme, a specific technique must be employed to generate \( S \) at a given level of \( c \), since the selection may depend on the variables' values. In both schemes, the coefficient \( c \) can, of course, be calculated a posteriori, that is, after the data has been generated, by analyzing the empirical distribution of the variables. However, it can be specified a priori only in the SCAR scheme.

In our experiments, we ensure a non-SCAR scenario, so we need to develop a technique to generate a surrogate variable. Our function is designed to create a new binary column, \( S \), within a dataframe by identifying and utilizing high-variance variables. The function selects the top \( n_{\text{vars}} \) columns with the highest variance and calculates intermediate variables that reflect cumulative sums influenced by the \( Y \) column. The linear approximation in the function is implemented to facilitate easier control over the coverage of the variable \( Y \) by the variable \( S \). By using a linear approximation, the function allows us to assign the desired \( c \). In practice, we usually obtain a dataset with a value close to the desired \( c \). We describe this in detail in the supplementary materials (Section 3. Description of the Function \texttt{non\_scar\_labelling.mvc}). In our experiments, we set \( n_{\text{vars}} = 1\)

This technique enables us to control the difficulty of the problem and simultaneously represents an attempt to model a complex disturbance of the true explanatory variable.
\subsection{Calibrations of Parameters and Experiment}
The naive logistic regression approach, the cluster method and the LassoJoint approach (strict and non-strict) have been employed. 
We deal with the problem of PU data classification under the non-SCAR assumption. From the above, completely labeled datasets, we select $c\cdot 100\%$ of labeled observations $S$, for (approximately) $c = 0.3; 0.5; 0.8$, and then, we randomly split these datasets into the training sample ($70\%$) and the test sample ($30\%$). As for the remaining values of our cluster cleaning method, we average the GLM coefficients $R=5$ times. We appropriately choose $q = 0.25, 0.5, 1$ from the known $S = 1$.
By applying the LassoJoint method in its first step, we use the Lasso method with tuning parameters $\lambda $, chosen either on the basis of the 10-fold cross-validation
scheme - in the first scenario (where lambda.min gives the minimum mean cross-validation error). In the second step, we apply the thresholded Lasso
design for $\delta =0.5\cdot\lambda$, with $\lambda$ selected in the first step (this choice of the parameters is one of the recommendation based on simulation study in \cite{F3}). Next, we determine the classification metrics by simulating from the Monte Carlo replications of our experiment. The number of repetitions depends on the size of the dataset. If our dataset has fewer than 100 columns or fewer than 10,000 rows, we will perform up to 100 experiments. However, if our dataset has 100 or more columns and 10,000 or more rows, we will limit the number of experiments to 10. This approach ensures that we are thorough in the case of small datasets, but efficient when dealing with larger datasets.

\begin{figure}[ht]\centering\includegraphics[scale = 0.41]{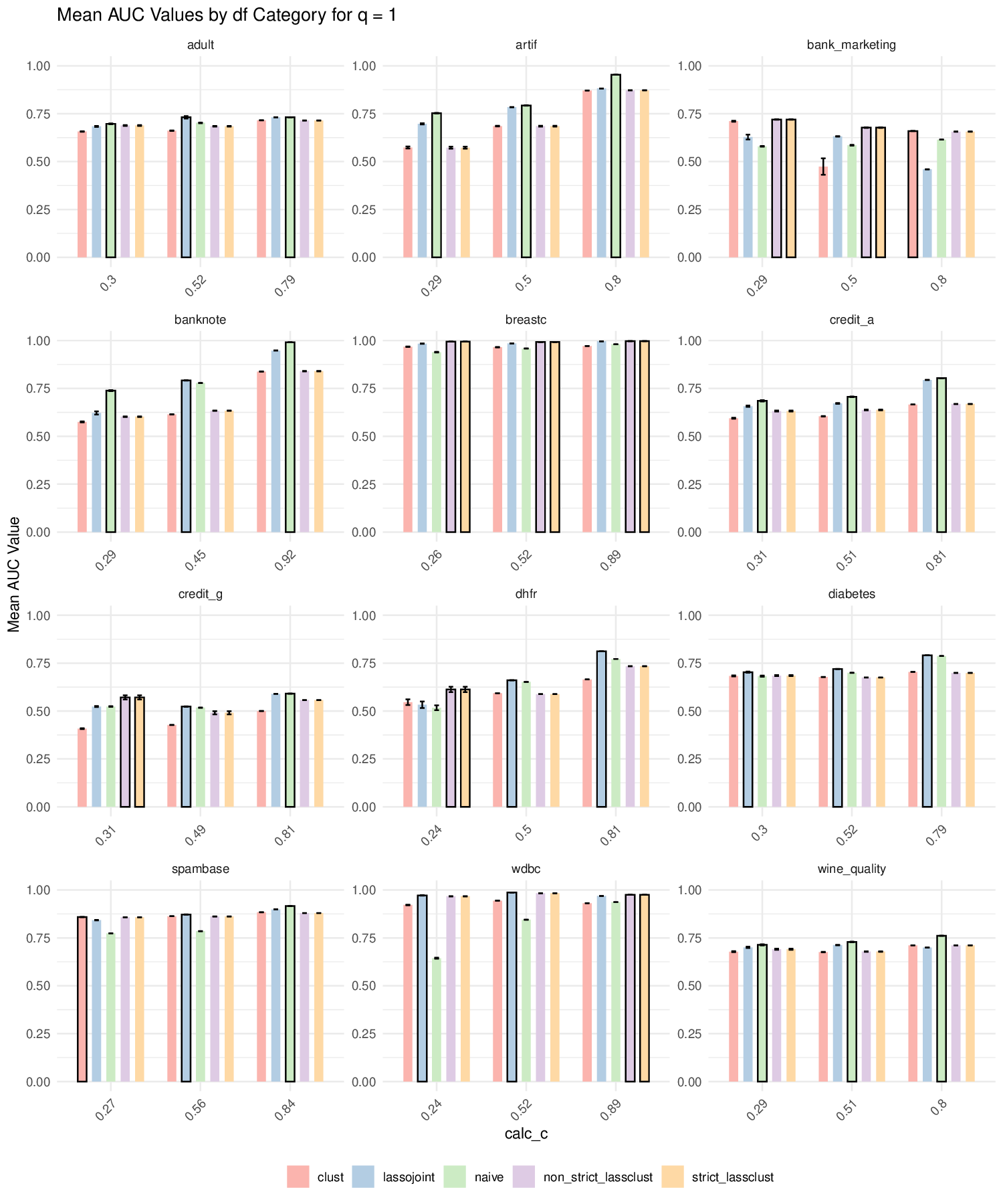}
\caption{Mean AUC Values for the methods for $q = 1$}
\label{fig:auc_values_q_1}
\end{figure}

\begin{table}[ht]
\centering
\tiny
\begin{tabularx}{\textwidth}{|l|X|X|X|X|X|X|X|X|X|X|X|X|X|X|X|}
\hline
\textbf{q} & \textbf{naive acc} & \textbf{clust acc} & \textbf{strict lassclust acc} & \textbf{non strict lassclust acc} & \textbf{lasso joint acc} & \textbf{naive f1} & \textbf{clust f1} & \textbf{strict lassclust f1} & \textbf{non strict lassclust f1} & \textbf{lasso joint f1} & \textbf{naive auc} & \textbf{clust auc} & \textbf{strict lassclust auc} & \textbf{non strict lassclust auc} & \textbf{lasso joint auc} \\ \hline
0.25 & \textbf{0.64 (0.16)} & \textbf{0.71 (0.16)} & 0.7 (0.15) & \textbf{0.71 (0.16)} & \textbf{0.53 (0.15)} & \textbf{0.63 (0.18)} & 0.54 (0.25) & 0.5 (0.25) & 0.54 (0.25) & 0.56 (0.14) & \textbf{0.77 (0.13)} & \textbf{0.73 (0.17)} & \textbf{0.75 (0.16)} & \textbf{0.75 (0.16)} & \textbf{0.79 (0.15)} \\ \hline
0.5 & \textbf{0.64 (0.16)} & \textbf{0.71 (0.16)} & 0.69 (0.15) & \textbf{0.71 (0.16)} & \textbf{0.53 (0.15)} & \textbf{0.63 (0.18)} & 0.54 (0.24) & 0.52 (0.23) & 0.54 (0.25) & 0.56 (0.14) & \textbf{0.77 (0.13)} & \textbf{0.73 (0.17)} & \textbf{0.75 (0.16)} & \textbf{0.75 (0.16)} & 0.78 (0.15) \\ \hline
1 & \textbf{0.64 (0.16)} & \textbf{0.71 (0.16)} & \textbf{0.71 (0.16)} & \textbf{0.71 (0.16)} & \textbf{0.53 (0.15)} & \textbf{0.63 (0.18)} & \textbf{0.56 (0.24)} & \textbf{0.56 (0.24)} & \textbf{0.56 (0.24)} & \textbf{0.57 (0.15)} & \textbf{0.77 (0.13)} & \textbf{0.73 (0.17)} & \textbf{0.75 (0.16)} & \textbf{0.75 (0.16)} & \textbf{0.79 (0.15)} \\ \hline

\end{tabularx}
\caption{Summary of Mean and Standard Deviation for Accuracy, F1, and AUC Metrics by q (Highest Mean in Bold); non-SCAR scheme}
\label{tab:summary_stats}
\end{table}

\subsection{Results}
\begin{figure}[h]
    \centering
    \includegraphics[width=\textwidth]{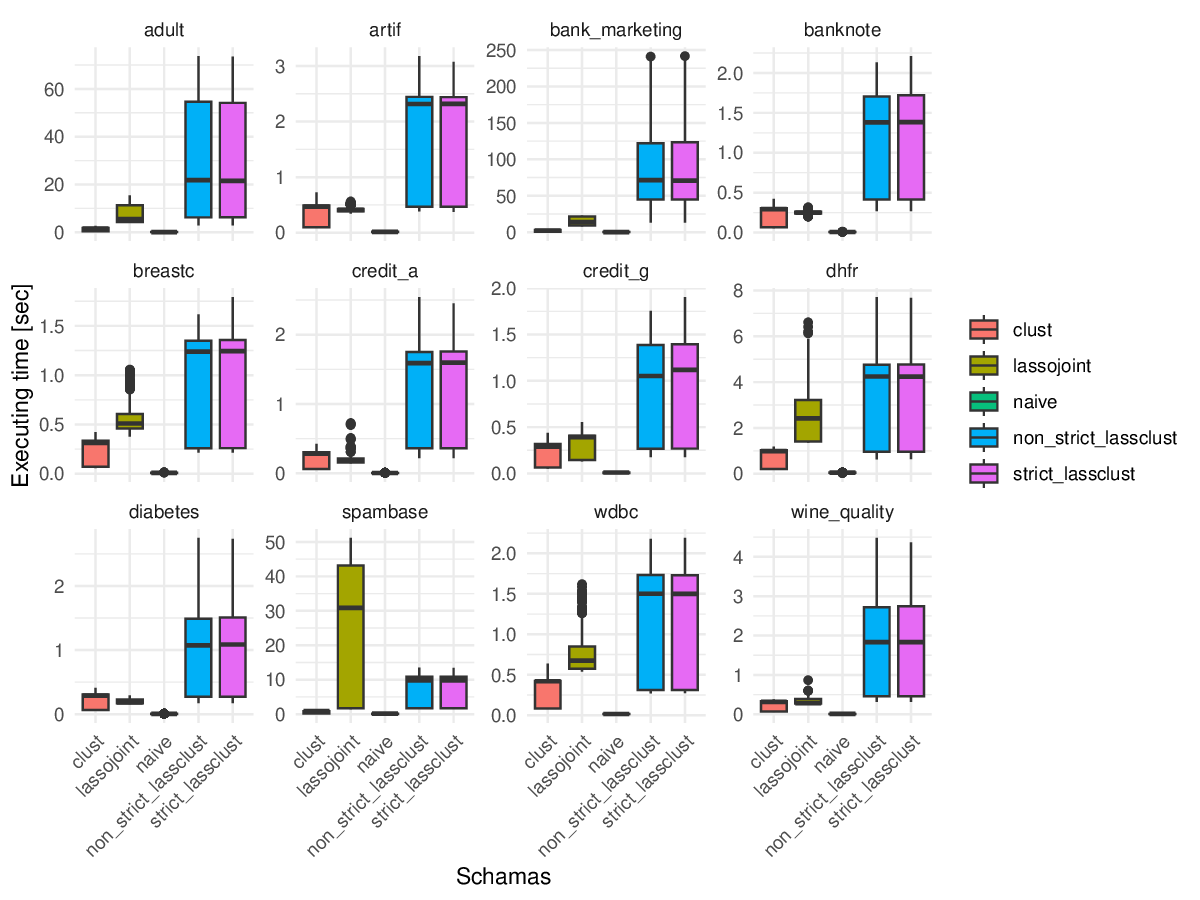}
    \caption{Boxplots of Executing time for the methods}
    \label{fig:boxplots}
\end{figure}

In the present simulation study, the non-SCAR and SCAR assumptions are considered. Now, we present results from a non-SCAR scenario.
The clust method has quite high accuracy on all sets except the banknote. The Lasso methods (strict and non strict) have similar accuracy to the clust method. The Lasso methods little improve the AUC for the breastc, credict\_a, credit\_g, dhfr, wdbc, wine\_quality. 
For $c>0.5$, the LassoJoint method performs quite well apart bank\_marketing (good accuracy, the AUC and F1 score). The LassoJoint method obtained very good accuracy for $c \geq 0.8$ for the sets: bank\_marketing, credit\_a and wine\_quality (see Figure \ref{fig:auc_values_q_1}).
There is no clear rule on which $q$ to choose. Based on Table \ref{tab:summary_stats} - the averaged statistics across all datasets and experiments suggest that the most likely value of $q$ should be large, around 1. However, in some cases, it might actually be small, around 0.25. This could depend on the specific characteristics of the variables in the dataset and this relationship is complex. 
Based on Figure \ref{fig:boxplots}, we have the shortest executing time for the naive method and the clust method, which is much shorter than methods using the Lasso method with cross-validation.
The results for all metrics are presented in our supplement materials (Section 1. Non-SCAR Scheme) \footnote{\url{https://github.com/kapacc/mdai25-clusters-scripts/blob/main/supplement.pdf}}. 
In the SCAR scenario, the best classification metrics are obtained for the LassoJoint method, which was expected since this method was invented with the SCAR assumption. A quite good result was obtained using the naive method and the clust method (except for the artif, banknotes and dhlr datasets). More information can be found in our supplementary materials (Section 2. SCAR Scheme).

\section{Conclusions}
In the considered model of the SCAR condition disturbance, clustering methods with the Lasso methods were quite effective in predicting the PU classification. In general, the use of Lasso regularization can improve the prediction. All the considered methods, with some exceptions in the case of the LassoJoint method, improve the quality measures of the classifiers as $c$ increases. As the value of $c$ increases, the PU classification problem becomes a classic full-label classification problem; hence, for $c$ close to 1, the naive method should gradually gain an advantage over other methods.
In situations where the SCAR condition is applicable, the LassoJoint algorithm, which was designed under this condition, performed nearly optimally. The clust algorithm also performed quite well, which allows us to assume that it can be used regardless of the SCAR condition. The presented work is a continuation and extension of the results from works \cite{F2}-\cite{F3} in the case when the SCAR condition is not met. 

Finally, it is worth noting that the first step of our procedure involving 2-means clustering can be applied not only to single-sample PU scenarios but also to case-control PU learning.
\\

\subsubsection{\discintname}
The authors declare that they have no known competing financial interests or personal relationships that could have appeared to
influence the work reported in this paper.
%
%
%
%

\end{document}